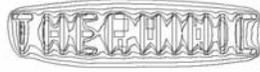

# ELECTRONICS COOLING FAN NOISE PREDICTION

*Antoine Dozolme[1], Hossam Metwally[2], Thierry Marchal[1].*

[1] Fluent Benelux, Avenue Pasteur 4, 1300 Wavre Belgium
[2] Fluent Inc., 1007 Church Street, Suite 250, Evanston, IL 60201, USA

## ABSTRACT

Using the finite volume CFD software FLUENT, one fan was studied at a given flow rate (1.5m³/min) for three different operational rotating speeds: 2,000, 2,350 and 2,700 rpm. The turbulent air flow analysis predicts the acoustic behavior of the fan. The best fan operating window, i.e. the one giving the best ratio between noise emissions and cooling performance, can then be determined. The broadband noise acoustic model is used. As the computation is steady state, a simple Multiple Reference Frame model (MRF, also known as stationary rotor approach) is used to represent the fan. This approach is able to capture the effects of the flow non-uniformity at the fan inlet together with their impact on the fan performance. Furthermore, it is not requiring a fan curve as an input to the model. When compared to the available catalog data the simulation results show promising qualitative agreement that may be used for fan design and selection purposes.

## 1. INTRODUCTION

In order to evacuate the quickly increasing heat generated by the always more powerful computers, new fans are constantly designed. The performance of these new designs in term of thermal evacuation is obvious. However, the quality of the fan is not restrained to its capability to evacuate heat through intense air flow anymore. The competitiveness of the end product, such as a laptop, now includes the technical performances of the computer but also the comfort it is maintaining to the end-users. Hence, the best designed fan has little value if it is too noisy. More and more fan designers are now including this noise dimension in the proper design criteria

Fans acoustics emissions are the main noise source in forced convection air cooled electronics systems. Improving the fan design for a given system will not only result in an increase in the amount of air flow it delivers, but may also lead to a reduction of noise emissions. A fan operating at a higher efficiency delivers more air for the same amount of torque; it is also less noisy within the same system. Fan noise is becoming an important issue especially with the increasing usage, and power, of portable electronics.

CFD analyses can help design the fan to improve both hydrodynamic and acoustical performance. In this work, steady state analysis of an axial fan is carried out at three different rpm's to estimate the broad band noise the fan emits. The surface acoustical power level is compared to available catalog data.

## 2. CFD MODELING

The GAMBIT package is used to generate the geometry and the mesh of the model. The flow is analyzed by solving the mass and momentum conservation equations. Momentum equation:

$$\frac{\partial}{\partial t}(\rho \vec{v}) + \nabla.(\rho \vec{v} \vec{v}) = -\nabla p + \rho \vec{g} + \vec{F} \qquad (1)$$

Continuity equation:

$$\frac{\partial \rho}{\partial t} + \nabla.(\rho \vec{v}) = S_m \qquad (2)$$

For a 3D model, the momentum equation is discretized as follow:

$$\sum_f^{N_{faces}} \rho_f \vec{v}_f \phi_f . \vec{A}_f = \sum_f^{N_{faces}} \Gamma_\phi (\nabla \phi)_n . \vec{A}_f + S_\phi V \qquad (3)$$

In this analysis, the model is isothermal. Turbulence is taken into account through the standard k-epsilon model. It is mathematically described by the following equations:

$$\frac{\partial}{\partial t}(\rho k) + \frac{\partial}{\partial x_i}(\rho k u_i) = \frac{\partial}{\partial x_j}\left[\left(\mu + \frac{\mu_t}{\sigma_k}\right)\frac{\partial k}{\partial x_j}\right] + G_k + G_b - \rho \varepsilon - Y_M + S_k \qquad (4)$$

$$\frac{\partial}{\partial t}(\rho \varepsilon) + \frac{\partial}{\partial x_i}(\rho \varepsilon u_i) = \frac{\partial}{\partial x_j}\left[\left(\mu + \frac{\mu_t}{\sigma_\varepsilon}\right)\frac{\partial \varepsilon}{\partial x_j}\right] + C_{1\varepsilon}\frac{\varepsilon}{k}(G_k + C_{3\varepsilon}G_b) - 2C_{2\varepsilon}\rho\frac{\varepsilon^2}{k} - R_\varepsilon + S_\varepsilon \qquad (5)$$

Where $G_k$ represents the generation of turbulence kinetic energy due to the mean velocity gradients, $G_b$ is the generation of turbulence kinetic energy due to buoyancy. $Y_M$ represents the contribution of the fluctuating dilatation in compressible turbulence to the overall dissipation rate.

*Antoine Dozolme[1], Hossam Metwally[2], Thierry Marchal[1].*

## *ELECTRONICS COOLING FAN NOISE PREDICTION*

The quantities $\sigma_k$ and $\sigma_\varepsilon$ are the inverse effective Prandtl numbers for k and $\varepsilon$, respectively. $S_k$ and $S_\varepsilon$ are user-defined source terms

The geometry consists of a tube. In the middle of the tube, a fan is explicitly modeled (figure 1). The mesh is made of 1,500,000 hexahedral elements. The inlet is represented in blue and the outlet is red. The walls are presented in light yellow with a mesh representation on them. The mesh is refined close to the fan in order to capture the flow field around it. The model is assumed isothermal, and the gravity is not taken into account.

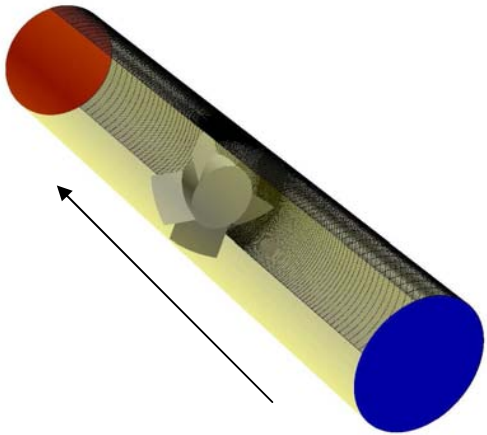

Fig 1: Geometry and mesh.

The inlet is defined as a constant velocity inlet. The air enters the domain at 2.5m/s, corresponding to the value of 1.5 m³/min. The outlet is specified as a free boundary condition. The entire domain walls except the fan walls have a zero velocity boundary condition. The fan is modeled using the MRF approach for each of the three velocities.

The standard k-epsilon model is used to describe the turbulence behavior. The value of k and epsilon will be used in the acoustics models to perform broadband acoustic calculation.

The air passing inside the fan has a density of 1.225kg/m³ and a viscosity of 1.78940.10⁻⁵Pa.s. The air compressibility is neglected.

### 3. BROADBAND MODEL

In many practical applications involving turbulent flows, noise does not have any distinct tones, and the sound energy is continuously distributed over a broad range of frequencies. In those situations involving broadband noise, statistical turbulence quantities readily computable from RANS equations (mean flow field, turbulent kinetic energy (k) and the dissipation rate ($\varepsilon$)) can be utilized, in conjunction with semi-empirical correlations and acoustic analogies, to shed some light on the source of broadband noise. Unlike the computationally expensive direct method (which involves computationally expensive detailed unsteady simulation) and the integral method, the broadband-noise-source models do not require transient solutions.

The broadband model is however limited to the broadband noise characteristics prediction and does not provide any tonal performance data

Proudman [2] ,using Lighthill's acoustic analogy, derived a formula for acoustic power generated by isotropic turbulence without mean flow. More recently, Lilley [3] re-derived the formula by accounting for the retarded time difference which was neglected in Proudman's original derivation. Both derivations yield acoustic power due to unit volume of isotropic turbulence.

The Proudman's formula gives an approximate measure of the local contribution to total acoustic power per unit volume in a given turbulence field. Proper caution, however, should be taken when interpreting the results in view of the assumptions made in the derivation, such as high Reynolds number, small Mach number, isotropy of turbulence, and zero mean motion.

### 4. MRF FORMULATION

The MRF model gives FLUENT the ability to model flows in an accelerating reference frame. In this situation, the acceleration of the coordinate system is included in the equations of motion describing the flow.

When problems such as axial fan are defined in a rotating reference frame, the rotating boundaries become stationary relative to the rotating frame, since they are moving at the same speed as the reference frame.

When the equations of motion are solved in a rotating frame of reference, the acceleration of the fluid is augmented by additional term that appears in the momentum equation. The left hand side of the momentum equation appears as follow for an inertial reference frame:

$$\frac{\partial}{\partial t}\left(\rho\vec{v}\right) + \nabla.\left(\rho\vec{v}\vec{v}\right) \qquad (6)$$

*Antoine Dozolme[1], Hossam Metwally[2], Thierry Marchal[1].*

### *ELECTRONICS COOLING FAN NOISE PREDICTION*

For flow in rotating domains, the equation for conservation of mass, or continuity equation can be written as follows:

$$\frac{\partial \rho}{\partial t} + \nabla . (\rho \vec{v}_r) = S_m \qquad (7)$$

## 5. RESULTS

The computational results are compared with available data from the manufacturer. The comparison is done for the static pressure rise and the noise prediction. The experimental data are obtained from [4].

Each computational run takes approximatively 3 hours on a 2 processors Pentium 4 machine running Linux.

Pressure rise is taken from a point in front of the fan to a point right after the fan. Table 1 shows the comparison of the results:

| rpm | Experimental | Computed |
|-----|-----|-----|
| 2,000 | 12.5 | 15.3 |
| 2,350 | 21.1 | 23 |
| 2,700 | 35.6 | 33.7 |

Table 1: Static pressure rise (Pa)

The computed results are less than 3Pa different with measurements what represents an average difference of about 10%. The faster the fan gets, the bigger the pressure rise. The next step is to compare the noise level. Table 2 shows the results.

| rpm | Experimental | Computed |
|-----|-----|-----|
| 2,000 | 38 | 48 |
| 2,350 | 42 | 55 |
| 2,700 | 47 | 62 |

Table 2: surface acoustic power level on the fan surface (dB)

The results are qualitatively comparable. The noise level is increasing when the rpm are increasing in both cases. The faster the fan gets, the noisier it is, as expected.
The noise distribution within the fan is then studied. For all three fans, the noise repartition is analyzed on the fan itself. Detailed analyses are described in section 5.2 to 5.4

### 5.1. Flow field study.

The air enters the domain at a constant velocity (2.5m/s). The flow direction is on the Z axis, from the lower Z values to the higher ones. When entering the fan, the air gets a swirl velocity from the fan, accelerating it (green to red on figure 2). After the fan, it clearly appears that the highest velocities remain on the larger radius of the cylindrical domain whereas the lowest velocities are found with the lower radius. On the blade, the larger velocities are found on the outside of the blades.

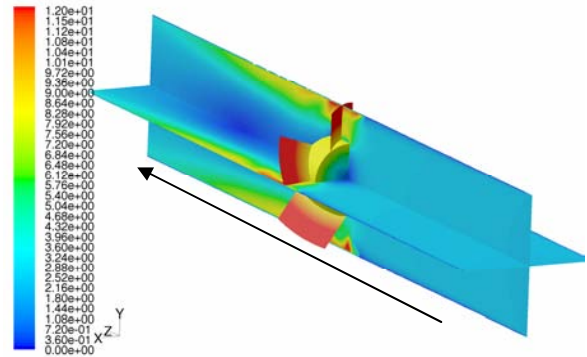

Fig 2: velocity magnitude (m/s) repartition in the computational domain for 2,700 rpm.

The pathlines representation illustrates the swirling flow induced by the fan. The air is going straight to the fan, after the device, the air is rotating until it gets out of the domain.

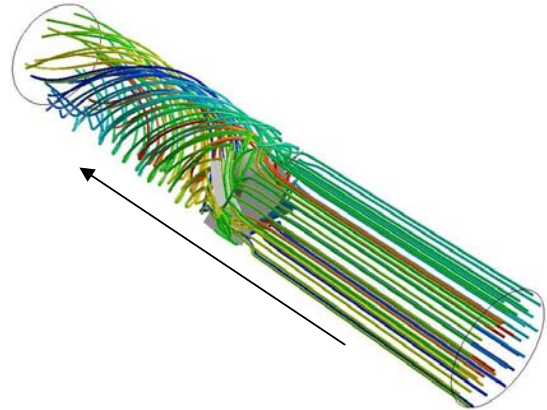

Fig 3: Path line inside the domain for 2,350 rpm. Path lines are colored by ID.

Inside the fan, the air is rotating. Figure 4 shows the air entering the fan with a quasi normal velocity, when the air is leaving the fan, it has been given a rotational movement.

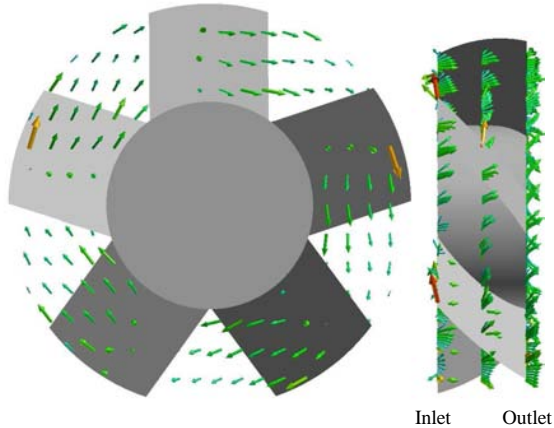

Inlet    Outlet

Fig 4: Velocity vector representation inside the fan for the 2,000 rpm case.

The pressure distribution on the fan for all three cases has the same repartition (figure 5, figure 6, figure 7). The pressure is increasing inside the fan. The lower pressures are located at the inlet of the fan whereas the higher pressures are located after the fan. At 2,000 rpm (figure 5), the pressure drop is 15.3Pa.

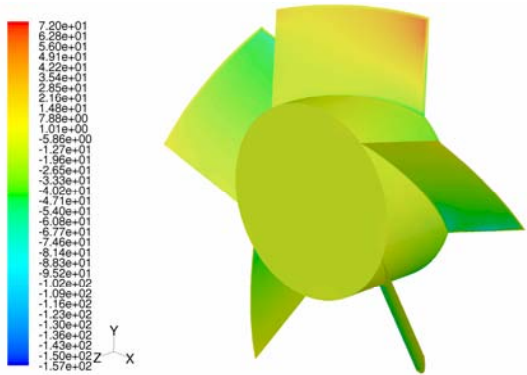

Fig 5: Static pressure (Pa) repartition on the fan at 2,000 rpm

At 2,350 rpm, negative static pressure locations are larger at the leading edges (figure 6). The maximum pressure has increased as well. The overall pressure drop as increased as well, up to 23 Pa.

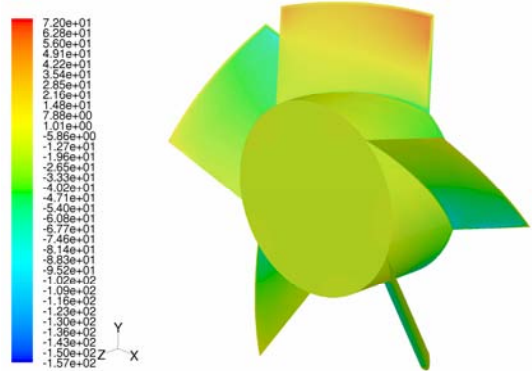

Fig 6: Static pressure (Pa) repartition on the fan at 2,350 rpm

At 2,700 rpm (figure 7), the negative static pressure zones have increased, as well as the maximum pressure on the blades. The overall pressure drop is now 33.7 Pa.

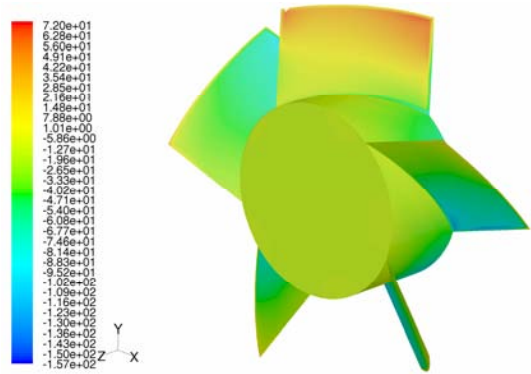

Fig 7: Static pressure (Pa) repartition on the fan at 2,700 rpm

These 3 cases illustrate that for a given flow rate, the static pressure difference is increasing with the rotational speed. This is what is expected from the fan curves from [4]. When the rotating speed increases, the fan curve is translated in a higher pressure drop region.

### 5.2. The 2,000 rpm case

The noise is the most important at the fan inlet. The noise is nearly non existent after the fan, and decreases inside the fan. At the leading edge of the blades, the noise is the largest.

*Antoine Dozolme[1], Hossam Metwally[2], Thierry Marchal[1].*

## ELECTRONICS COOLING FAN NOISE PREDICTION

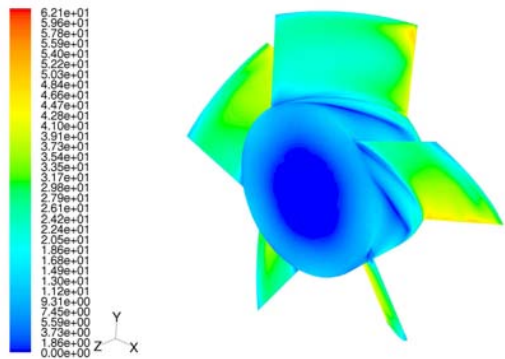

Fig 8: acoustic level (dB) repartition on the fan at 2,000 rpm

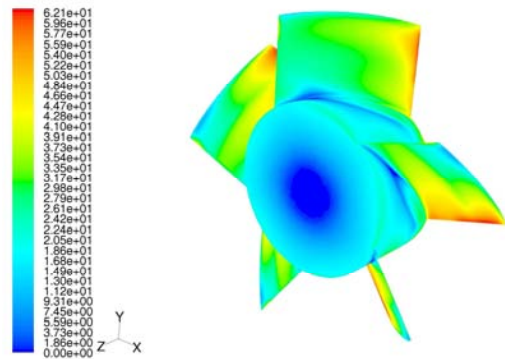

Fig 10: acoustic level (dB) repartition on the fan at 2,700 rpm

The sound level repartition follows the same contours as the pressure, but it is the larger where the pressure is the lower.

### 5.3. The 2,350 rpm case

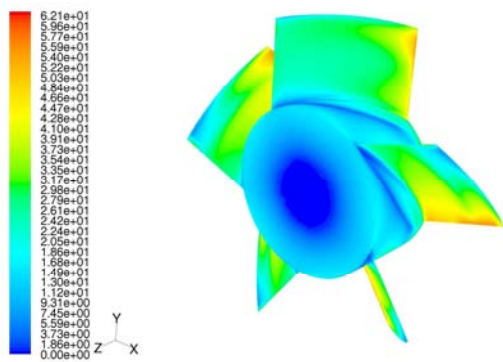

Fig 9: acoustic level (dB) repartition on the fan at 2,350 rpm

The same behaviors as for the 2,000 rpm case are observed. There is a slight increase of the noise level and pressure on the blades. The noise repartition remains the same, but at a higher level.

Most of the noise is observed on the negative static pressure zones on the blades. Interestingly, the zones with no noise on the fan hub (blue part) are occurring where the pressure is zero.

### 5.4. The 2,700 rpm case

The noise repartition remains the same as for the 2,350 rpm case on the fan body. The slight difference is a decrease of the silent part in front of the fan. The global noise level has slightly increased as well.

Increasing the fan rotating speed increases the noise level for a given mass flow rate.

## 6. CONCLUSIONS

FLUENT can predict the noise repartition inside a fan for several flow regime. The comparisons with experimental data for both global noise level and static pressure drop are comparable with the computed results. The broadband model gives accurate results for this study , for a limited CPU time, due to its physical formulation (steady state, standard k-epsilon turbulence model).